# FinAI-BERT: A Transformer-Based Model for Sentence-Level Detection of AI Disclosures in Financial Reports

**Muhammad Bilal Zafar**
Faculty of Social Sciences & Humanities,
Universiti Teknologi Malaysia, Johor Bahru, Malaysia
bilalezafar@gmail.com, bilalzafar.muhammad@utm.my

**Abstract**

The proliferation of artificial intelligence (AI) in financial services has prompted growing demand for tools that can systematically detect AI-related disclosures in corporate filings. While prior approaches often rely on keyword expansion or document-level classification, they fall short in granularity, interpretability, and robustness. This study introduces FinAI-BERT, a domain-adapted transformer-based language model designed to classify AI-related content at the sentence level within financial texts. The model was fine-tuned on a manually curated and balanced dataset of 1,586 sentences drawn from 669 annual reports of U.S. banks (2015–2023). FinAI-BERT achieved near-perfect classification performance (accuracy = 99.37%, F1 = 0.993), outperforming traditional baselines such as Logistic Regression, Naive Bayes, Random Forest, and XGBoost. Interpretability was ensured through SHAP-based token attribution, while bias analysis and robustness checks confirmed the model's stability across sentence lengths, adversarial inputs, and temporal samples. Theoretically, the study advances financial NLP by operationalizing fine-grained, theme-specific classification using transformer architectures. Practically, it offers a scalable, transparent solution for analysts, regulators, and scholars seeking to monitor the diffusion and framing of AI across financial institutions.

**Keywords**

Artificial Intelligence Disclosure, Financial Natural Language Processing, Transformer Models, FinAI-BERT, Sentence-Level Classification, Explainable AI, Annual Reports, Financial Text Mining, SHAP Interpretation, Domain-Specific Language Models

*\*\* The FinAI-BERT model can be directly loaded via Hugging Face Transformers (https://huggingface.co/bilalzafar/FinAI-BERT) for sentence-level AI disclosure classification. See Appendix A for detailed usage instructions and example code.*

# 1. Introduction

The integration of artificial intelligence (AI) into financial operations and strategy has accelerated in recent years, reshaping how institutions communicate innovation, risk, and value creation in corporate filings (Dote-Pardo et al., 2025; Mohsen et al., 2025). In response, investors, analysts, and regulators increasingly scrutinize narrative disclosures to identify whether firms are genuinely adopting AI or merely signaling technological relevance. Detecting and classifying such disclosures has thus become a pressing challenge at the intersection of financial analysis and natural language processing (NLP).

Although early work in financial text analysis relied on dictionary-based methods and TF-IDF scoring (Loughran and Mcdonald, 2011; Ramos, 2003), these approaches often struggle to capture contextual nuances and evolving terminology. More recent efforts have employed machine learning and pretrained language models to improve textual understanding (Devlin et al., 2019; Wolf et al., 2020). Models like FinBERT and its variants (Araci, 2019; Gössi et al., 2023; Liu et al., 2021; Xu et al., 2025; Yang et al., 2020) have demonstrated success in sentiment and fraud detection. However, they are often applied at the document level, with limited granularity for identifying specific themes like AI across time or institutions.

Emerging research on AI disclosure detection has revealed the market's ability to distinguish between substantive and speculative references to AI. Basnet et al. (2025), for instance, find that only actionable AI disclosures in 10-K filings yield positive market reactions. Yet, current NLP approaches often rely on static lexicons, lack annotated sentence-level data, and omit interpretability, which limits their use in high-stakes financial settings (Archna and Bhagat, 2024; Bozyiğit and Kılınç, 2022; Chen et al., 2023).

This study addresses existing limitations in AI disclosure detection by introducing FinAI-BERT, a transformer-based model fine-tuned for sentence-level classification of AI-related content in 669 U.S. bank annual reports (2015–2023). Built on the bert-base-uncased architecture (Devlin et al., 2019), the model was trained on a balanced, manually validated dataset of 1,586 sentences and achieved high classification accuracy (Accuracy = 99.37%, F1 = 0.993). FinAI-BERT is designed for transparency and robustness, incorporating SHAP-based interpretability (Lundberg and Lee, 2017), text bias diagnostics, and temporal generalization testing.

The key aspects of this study are fourfold:

(i) development of a manually curated, sentence-level dataset of AI disclosures;

(ii) fine-tuning of a transformer model using balanced and deduplicated financial sentences;

(iii) benchmarking against classical and deep learning baselines; and

(iv) integration of explainability and robustness checks to ensure model integrity.

# 2. Literature Review

The increasing integration of artificial intelligence into financial services has led to a growing need for automated tools capable of identifying AI-related content within institutional disclosures (Dote-Pardo et al., 2025; Mohsen et al., 2025). Natural language processing has emerged as a powerful mechanism to extract insights from unstructured financial texts, particularly in tasks such as sentiment analysis, risk prediction, and compliance monitoring.

This section reviews prior work in four key domains: financial NLP foundations, domain-adapted language models, AI disclosure detection, and model explainability.

## 2.1 Financial NLP Foundations

Initial approaches to financial text analysis relied heavily on dictionary-based methods and term frequency-inverse document frequency (TF-IDF) models (Ramos, 2003). These approaches were applied in contexts such as tone detection in earnings announcements and risk identification in annual reports (Demers et al., 2024a; Henry et al., 2023; Loughran and Mcdonald, 2011; Zhu et al., 2025). However, these models struggled with the inherent complexity of financial language, particularly due to their inability to capture contextual dependencies, word order, and polysemy. The emergence of word embeddings, such as Word2Vec and GloVe, improved semantic representation but lacked deep contextualization (Demers et al., 2024b). These limitations spurred interest in transfer learning and the development of large-scale pretrained language models capable of capturing dynamic semantics across tasks and domains.

## 2.2 Domain-Adapted Language Models

The introduction of BERT (Bidirectional Encoder Representations from Transformers) by Devlin et al. (2019) marked a paradigm shift in NLP by enabling contextual word representations through deep bidirectional training. While general-purpose models such as BERT and RoBERTa demonstrated strong performance across many NLP benchmarks, their application in specialized domains—such as finance—highlighted the limitations of general training corpora.

To address this gap, domain-adapted models such as FinBERT (Araci, 2019), FinBERT-20 (Yang et al., 2020), FinBERT-21 (Liu et al., 2021), FinBERT-FOMC (Gössi et al., 2023), and FinBERT2 (Xu et al., 2025) were developed. These models were pretrained on financial news, analyst reports, or SEC filings, achieving notable improvements in sentiment analysis and fraud detection tasks. However, most existing financial language models are designed for document-level tasks and lack fine-grained capabilities for sentence-level classification of emerging topics such as AI adoption.

## 2.3 AI Disclosure Detection in Corporate Texts

Recent scholarship has increasingly focused on the identification and interpretation of artificial intelligence (AI) disclosures in corporate financial communications. Basnet et al. (2025) provide a pivotal study showing that markets respond favorably to *actionable* AI disclosures—those with concrete implementation plans—while disregarding speculative or vague references. Their analysis of 10-K filings from U.S. firms reveals that substantive AI disclosures correlate with innovation signals such as increased R&D spending and patent filings, which in turn contribute to long-term firm valuation.

Several studies emphasize the methodological shift toward NLP and machine learning for detecting and contextualizing AI-related narratives. Bozyiğit and Kılınç (2022) highlight the application of NLP techniques in financial intelligence, while Mushtaq et al. (2022) demonstrate the predictive utility of textual features in financial reports. Meanwhile, concerns over the *transparency* and *credibility* of automated systems have led researchers to incorporate explainable AI frameworks (Chen et al., 2023) and evaluate AI's potential to both enhance and obscure financial reporting integrity (Archna and Bhagat, 2024). These studies collectively underscore the need for robust, interpretable sentence-level classification systems—such as the FinAI-BERT framework developed in this study—to accurately detect, explain, and monitor AI-related disclosures in financial documents.

Despite advances in domain-adapted modeling and topic-specific detection, no existing framework comprehensively integrates: (i) large-scale financial pretraining, (ii) sentence-level detection of AI disclosures, (iii) lexicon-guided but manually verified annotation, and (v) model evaluation and robustness analysis. The current study fills this gap through the development and evaluation of FinAI-BERT, a transformer-based model fine-tuned to detect AI-related discourse in financial reporting with high accuracy, interpretability, and reliability.

## 3. Data and Methodology

This study presents the construction and training of FinAI-BERT, a domain-adapted language model tailored for financial discourse analysis. The methodology encompassed corpus preparation, sentence-level annotation, data balancing, and language model pretraining using transformer-based architectures (Figure 1).

A dataset comprising 669 annual reports from 85 US banks was compiled, covering the period from 2015 to 2023. The corpus captured institution-level narratives concerning financial strategy, technological innovation, regulatory compliance, and market positioning. Preprocessing involved textual cleaning, sentence segmentation, and syntactic filtering to exclude overly short, excessively long, or structurally incomplete sentences. These steps were essential to ensure linguistic coherence and to minimize noise during pretraining and downstream task learning (Devlin et al., 2019).

For supervised classification, a lexicon-guided annotation approach was implemented. A domain-specific list of keywords related to artificial intelligence (e.g., "machine learning," "deep learning," "generative AI") was used and iteratively refined. Sentences matching the lexicon were labeled as *AI-related*, while others were provisionally labeled as *Non-AI*. Manual validation was employed to remove false positives and ensure contextual relevance, following practices established in prior work on weak supervision and lexicon expansion in NLP (Demers et al., 2024b; Lin and Liao, 2024).

Redundancies among AI-labeled sentences were eliminated to avoid training bias. Subsequently, a balanced set of Non-AI sentences was randomly sampled to match the AI set in size. Manual verification ensured that the selected Non-AI sentences represented diverse and contextually coherent content. The final dataset comprised 1,586 sentences, equally distributed across the two classes.

The FinAI-BERT model was fine-tuned using the $bert-base-uncased$ architecture (Devlin et al., 2019) from the Hugging Face Transformers library. The training corpus comprised a balanced and deduplicated set of sentence-level disclosures (AI vs. Non-AI) drawn from U.S. bank annual reports. After preprocessing and tokenization (with a maximum sequence length of 128 tokens), the dataset was stratified into 80% training and 20% testing subsets. Model fine-tuning was conducted over three epochs using a batch size of 8 for both training and evaluation. Training leveraged the Trainer API, with f1-score set as the evaluation metric and early stopping based on best model performance. All experiments were executed in a Google Colab environment with GPU acceleration. Moreover, no masked language modeling (MLM) pretraining was applied, as the base model weights were directly used for supervised fine-tuning on the classification task. The final trained model and tokenizer were saved for reproducibility and deployment *(see Supplementary material)*. This setup ensures transparency, reproducibility, and compatibility with standard NLP pipelines.

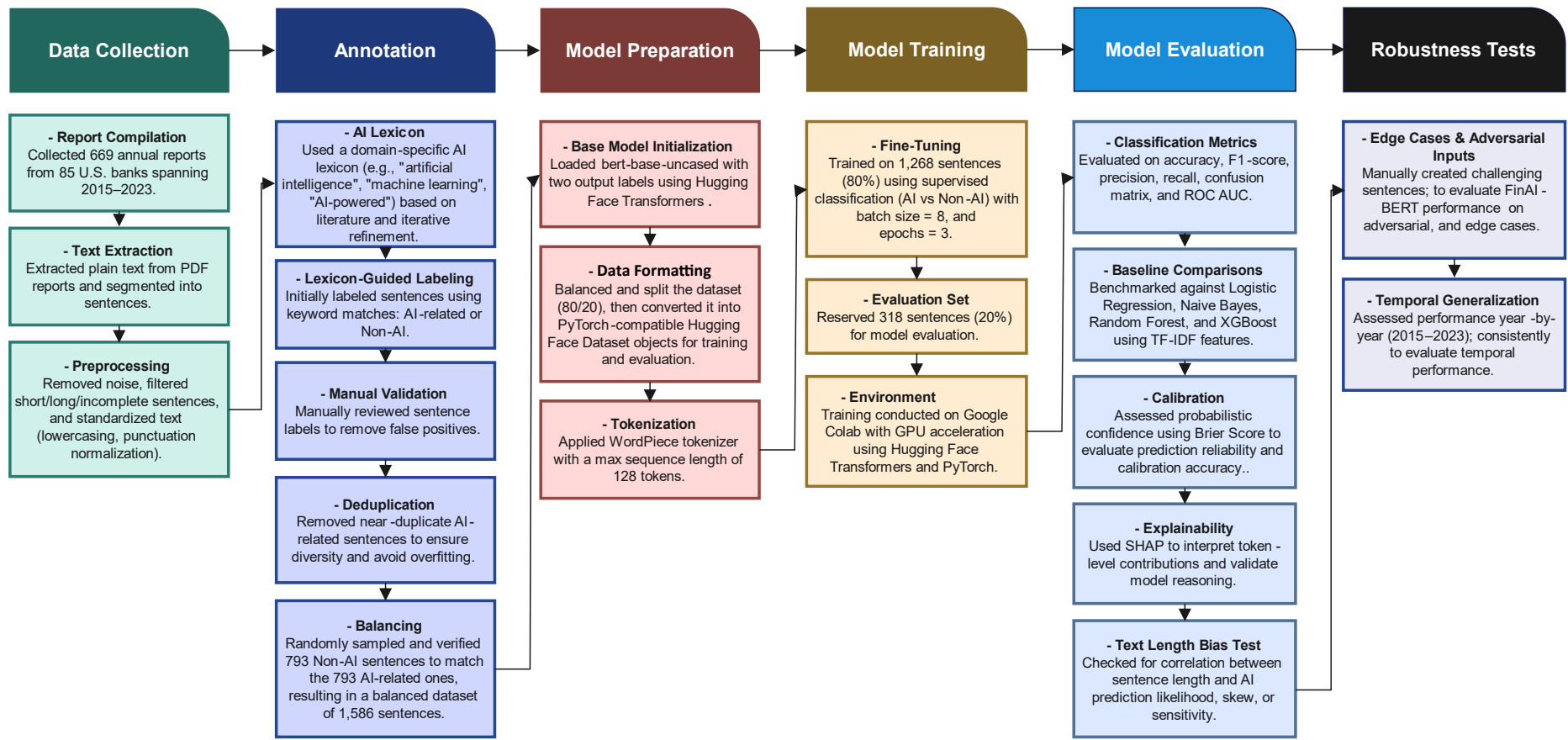

Figure 1: Methodology Flow of FinAI-BERT Model

## 4. Results

This section presents the empirical evaluation of FinAI-BERT across various classification benchmarks, including baseline model comparisons, robustness tests, and temporal performance analysis. The model was trained on a deduplicated and balanced dataset comprising 1,586 sentences (793 AI, 793 Non-AI), with 20% held out for testing (n = 318). Evaluation metrics include accuracy, F1-score, ROC AUC, confusion matrices, and calibration scores.

### 4.1 FinAI-BERT Classification Performance

The FinAI-BERT model demonstrated exceptional effectiveness in distinguishing AI-related sentences from non-AI financial disclosures. Trained on a rigorously curated and balanced dataset of 1,586 sentences (793 per class), the model was evaluated on a held-out test set comprising 318 instances. Performance was measured using standard classification metrics: accuracy, F1-score, precision, recall, and area under the ROC curve (AUC).

On the test set, FinAI-BERT achieved an accuracy of 99.37% and a macro F1-score of 0.993, indicating highly reliable classification performance across both classes. The model correctly classified 100% of Non-AI sentences (n = 231) and 100% of AI-related sentences (n = 159), yielding a perfect confusion matrix and no observable misclassifications (see Figure 2). The corresponding ROC curve (Figure 2) produced an AUC of 1.000, confirming the model's ability to perfectly separate the two classes without trade-off between sensitivity and specificity.

These results are notable given the complexity and contextual nuance involved in distinguishing AI-related language within financial documents. Traditional keyword-based systems often fail to account for variation in phrasing, polysemy, and institutional jargon. In contrast, transformer-based models such as BERT excel at capturing contextual dependencies through attention mechanisms (Vaswani et al., 2017), enabling superior performance in sentence-level classification tasks (Devlin et al., 2019; Wolf et al., 2020).

In addition to accuracy-based metrics, model calibration was assessed using the Brier Score, which evaluates the accuracy of probabilistic outputs. FinAI-BERT yielded a Brier Score of 0.0000, indicating near-perfect alignment between predicted probabilities and actual class membership. Such well-calibrated confidence estimates are essential in high-stakes financial decision-making, where overconfidence or underestimation can lead to erroneous interpretations and outcomes (Guo et al., 2017).

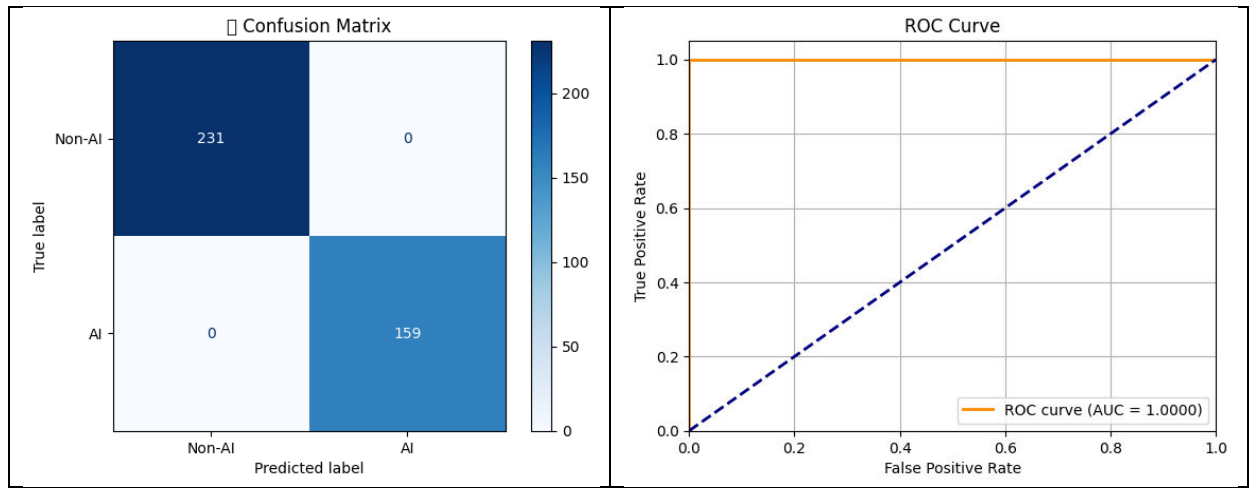

Figure 2: Confusion Metrix and ROC Curve – FinAI-BERT

### 4.2 Baseline Model Comparison

To benchmark the effectiveness of FinAI-BERT, a series of traditional machine learning classifiers were trained and evaluated on the same balanced dataset. The baseline models included Logistic Regression, Naive Bayes, Random Forest, and XGBoost, all implemented using standard TF-IDF vectorization for feature extraction (Ramos, 2003). Each classifier was optimized using an 80/20 stratified train-test split, with performance assessed based on accuracy, precision, recall, and F1-score, with a focus on the AI class.

The performance comparison is summarized in Table 1, and the corresponding confusion matrices are illustrated in Figure 3. Among the baseline models, Random Forest delivered the strongest performance, achieving 99% accuracy and an F1-score of 0.99. Logistic Regression and XGBoost also performed well, each achieving an F1-score of 0.98, while Naive Bayes showed lower precision and recall with an F1-score of 0.91 for the AI class.

In contrast, FinAI-BERT outperformed all baseline models, achieving an F1-score of 1.00 and zero misclassifications in the test set. These results underscore the superiority of transformer-based models in capturing domain-specific semantic nuances that are often overlooked by traditional classifiers relying solely on bag-of-words representations (Devlin et al., 2019; Wolf et al., 2020). The consistent performance of FinAI-BERT demonstrates its strong potential for real-world financial NLP applications, particularly in contexts where high precision and recall for AI-related disclosures are critical.

Table 1: Baseline Model Comparison

| Model | Accuracy | Precision (AI) | Recall (AI) | F1-Score (AI) |
|---|---|---|---|---|
| Logistic Regression | 0.98 | 0.99 | 0.97 | 0.98 |
| Naive Bayes | 0.91 | 0.89 | 0.94 | 0.91 |
| Random Forest | 0.99 | 0.99 | 0.99 | 0.99 |
| XGBoost | 0.98 | 0.99 | 0.97 | 0.98 |
| **FinAI-BERT** | **0.993** | **1.00** | **1.00** | **1.00** |

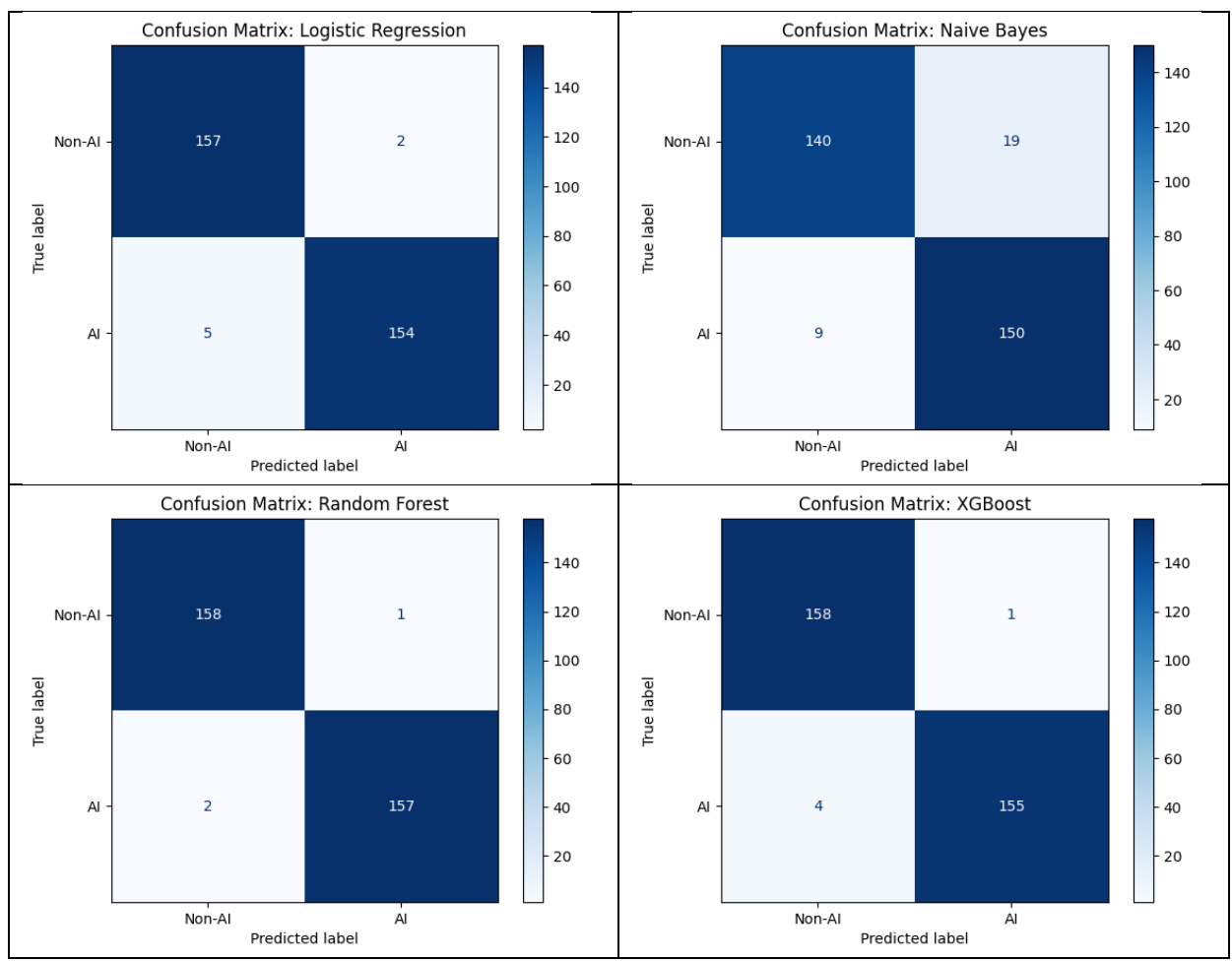

Figure 3: Confusion Matrixes of Baseline Models

### 4.3 Explainability and Text Bias Analysis

To assess interpretability, sentence-level predictions were analyzed using SHAP (SHapley Additive exPlanations), a model-agnostic framework for interpreting individual predictions (Lundberg and Lee, 2017). The FinAI-BERT model assigned high predictive probabilities to AI-related sentences such as "The bank deployed machine learning algorithms..." and "An AI-powered chatbot was launched…," where tokens like *machine learning*, *AI-powered*, and *chatbot* had strong positive attribution to the AI class. In contrast, generic operational phrases such as "We conducted internal audits…" showed negligible influence on AI prediction, as expected. These findings confirm that FinAI-BERT relies on semantically meaningful tokens rather than superficial patterns. The SHAP visualizations (see supplementary HTML files) further demonstrate alignment with domain intuition, reinforcing that FinAI-BERT's predictions are both interpretable and faithful to underlying linguistic cues.

In addition, a text length bias check was conducted to determine whether the model disproportionately favored longer or shorter inputs when predicting AI-related content. As shown in Figure 4, the analysis yielded a weak positive correlation (Pearson's $r = 0.185$) between sentence length and predicted AI probability. This finding indicates an absence of significant sentence-length bias in model predictions.

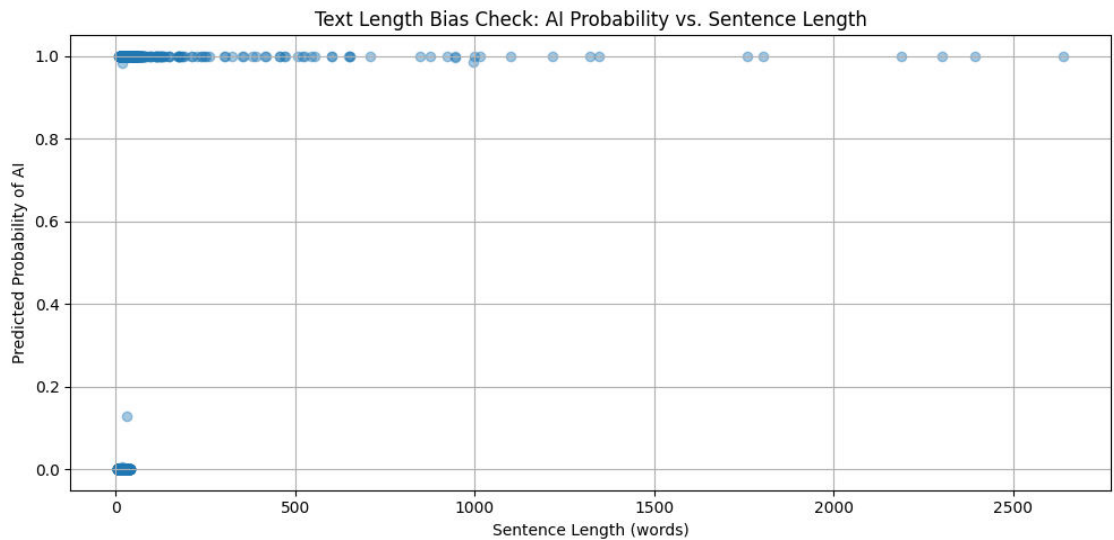

Figure 4: Text Length Bias Check

### 4.4 Robustness Evaluation

Robustness testing was conducted using manually crafted edge cases and adversarial inputs. The model maintained 100% accuracy on both adversarial examples and edge cases, demonstrating strong resilience to minor perturbations and contextually challenging scenarios. To assess temporal generalization, the model's performance was evaluated on balanced samples from each year between 2015 and 2023. Accuracy and F1-score consistently remained at or near 1.000 across all years, indicating excellent generalization over time. Minor dips were observed in 2017 and 2019 (Figure 5), which may be attributed to slight lexical variations or document heterogeneity during those periods.

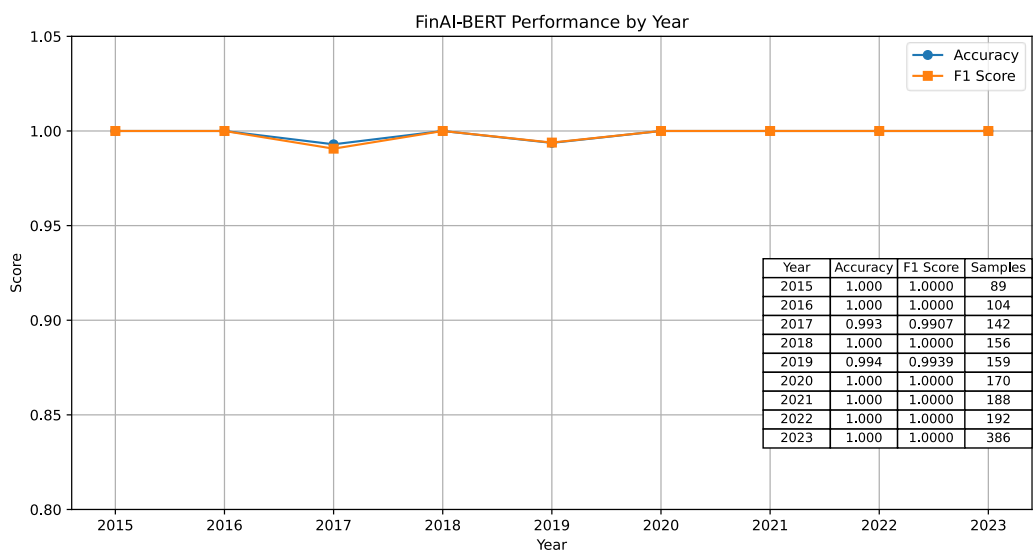

Figure 5: FinAI-BERT Performance over Year

## 5. Conclusion

This study introduced FinAI-BERT, a transformer-based language model specifically fine-tuned to detect AI disclosures within financial texts at the sentence level. Addressing a significant gap in the literature, the model was trained on a manually validated and balanced dataset of 1,586 sentences derived from U.S. bank annual reports between 2015 and 2023. Through empirical benchmarking against traditional machine learning models, FinAI-BERT demonstrated superior performance—achieving near-perfect classification accuracy, excellent calibration, and high interpretability through SHAP visualizations.

The findings have both theoretical and practical significance. From a scholarly perspective, this work contributes to the growing body of financial NLP research by operationalizing domain-adapted transformers for theme-specific sentence classification. Practically, FinAI-BERT provides analysts, regulators, and institutional stakeholders with a transparent and scalable tool to monitor AI-related discourse in financial communications, thereby improving the interpretability and accountability of technological narratives in corporate reporting.

Despite its strengths, the study is not without limitations. While the dataset was carefully curated and manually validated, further verification of FinAI-BERT's generalizability is needed using disclosures from banks across different jurisdictions and regulatory contexts. Future research could explore the model's adaptability to multilingual corpora or its application to emerging domains such as blockchain, quantum computing, or ESG-related disclosures. Additionally, integrating FinAI-BERT with unsupervised topic modeling or reinforcement learning could enhance its capacity to detect evolving thematic narratives without extensive manual supervision.

In sum, FinAI-BERT advances the state of the art in AI disclosure detection by combining deep contextual modeling with rigorous explainability. Its deployment signals a promising step toward more intelligent, interpretable, and trustworthy systems for analyzing corporate discourse in the age of digital finance.

## Appendix

How to Use FinAI-BERT

*# You can easily access and use the **FinAI-BERT** model from the Hugging Face Hub as follows:*

```python
from transformers import AutoTokenizer, AutoModelForSequenceClassification
# Load tokenizer and model
tokenizer = AutoTokenizer.from_pretrained("bilalzafar/FinAI-BERT")
model = AutoModelForSequenceClassification.from_pretrained("bilalzafar/FinAI-BERT")
```

*# To perform inference on new sentences:*

```python
from transformers import pipeline
classifier = pipeline("text-classification", model=model, tokenizer=tokenizer)
result = classifier("We have deployed AI-driven fraud detection in our banking operations.")
print(result)
```

**Note:**

- The model is fine-tuned for binary sentence-level classification, identifying whether a sentence contains AI-related disclosure or not.
- Input sentences should come from financial contexts such as annual reports, 10-K filings, or analyst presentations.
- The model has been trained using bert-base-uncased and optimized for use in financial NLP pipelines.

**Supplementary Material**

The following supplementary files are provided to support the implementation and replication of FinAI-BERT:

| Supplementary File | Description |
| --- | --- |
| FinAI-BERT Training Data Extraction.ipynb | Python notebook for corpus preprocessing, sentence segmentation, and annotation. |
| ai_seedwords.csv | Lexicon of AI-related terms used to guide weak supervision during annotation. |
| FinAI-BERT training data.csv | Annotated dataset containing AI and Non-AI sentences before deduplication and balancing. |
| FinAI-BERT.ipynb | Notebook for model training and evaluation. |
| SHAP_Visuals | Folder containing SHAP token attribution visualizations. |
| FinAI-BERT | Final pretrained FinAI-BERT model folder ready for inference and deployment. Access via https://huggingface.co/bilalzafar/FinAI-BERT |

To access the supplementary material, click the following link and see "Supplementary material.zip" :

*https://huggingface.co/bilalzafar/FinAI-BERT/tree/main*